\documentclass[fleqn,usenatbib]{rasti}

\usepackage{newtxtext,newtxmath}

\usepackage[T1]{fontenc}

\DeclareRobustCommand{\VAN}[3]{#2}
\let\VANthebibliography\thebibliography
\def\thebibliography{\DeclareRobustCommand{\VAN}[3]{##3}\VANthebibliography}

\usepackage{graphicx}	
\usepackage{amsmath}	
\usepackage{hyperref}

\title[K dwarf starspot hunters]{K-dwarf starspot hunters: how to directly characterize spot and faculae properties with UV-IR panchromatic transit spectra}

\author[J. K. Barstow \& C. Cullen et al.]{Joanna K. Barstow$^{1}$\thanks{E-mail: Jo.Barstow@open.ac.uk (JKB)}
\& Caitlyn Cullen$^{1}$\thanks{E-mail: Caitlyn.Cullen@open.ac.uk (CC)},
Yvonne Unruh$^{2}$, 
Hannah Wakeford$^{3}$,
Eva-Maria Ahrer$^{4}$, \newauthor
Lili Alderson$^{5}$,
Charlotte Fairman$^{3}$, 
Ulrich Kolb,$^{1}$
Nikole Lewis$^{5}$ \&
Cinta Vidante$^{4}$ 
\\
$^{1}$School of Physical Sciences, The Open University, Walton Hall, Milton Keynes, MK7 6AA, UK\\
$^{2}$Department of Physics, Imperial College London, Prince Consort Road, London, SW7 2AZ, UK\\
$^{3}$School of Physics, HH Wills Laboratory, Tyndall Avenue, Bristol, BS8 1TL \\
$^{4}$Max Planck Institute for Astronomy (MPIA), K\"{o}nigstuhl 17, 69117 Heidelberg, Germany \\
$^{5}$Department of Astronomy, Cornell University, 122 Sciences Drive, Ithaca, NY 14853, USA \\
}

\date{Accepted XXX. Received YYY; in original form ZZZ}

\pubyear{\the\year{}}

\begin{document}
\label{firstpage}
\pagerange{\pageref{firstpage}--\pageref{lastpage}}
 \maketitle

\begin{abstract}
Transmission spectroscopy of exoplanets provides an important window into their atmospheric composition, structure and dynamics, especially in the JWST era. As the quality of the observational data and therefore the precision of our constraints improves, we become increasingly sensitive to sources of systematic bias in our models and observations. One significant example of this is the Transit Light Source Effect, in which heterogeneous features in the stellar photosphere -- spots and faculae -- imprint additional spectral signatures onto exoplanet transmission spectra. The effects of this stellar contamination must be accounted for or removed in order to accurately recover the planet's atmospheric properties. Including parameterized models of spots and faculae within spectral retrieval analysis is increasingly adopted as a solution to this problem, but is limited by the accuracy of models of stellar spectra. In this paper, we outline an observational strategy combining simultaneous data from JWST and the Hubble Space Telescope that would enable us to further constrain these models. We perform synthetic retrievals to demonstrate how we could constrain starspot and faculae parameters; we test the impact of using different model spectra in the retrieval; and we consider the impact of variable stellar activity on coadding transits to achieve the desired signal-to-noise. We find that ultraviolet and optical wavelengths are key for breaking degeneracies between stellar contamination and planetary parameters. 
\end{abstract}




\section{Introduction}
The surfaces of most low-mass main-sequence stars are marked by spots (dark regions), faculae (bright regions), or a combination of the two, much like the surface of the Sun \citep{berdyugina2005,reinhold2019}. These features arise due to magnetic activity at the star's surface, with spots marking areas where magnetic flux inhibits convection, and faculae forming between convective granules. The average number and variability of these features generally correlates with other metrics for stellar activity \citep[e.g.][]{sowmya2023}, such as the log($R'_{HK}$) index \citep{noyes1984} or S index \citep{vaughan1978} which are measures of chromospheric Ca II H and K emission, although specific chromospheric features may not be colocated with photospheric heterogeneities \citep{morris2018}. Starspot activity is generally greater for later spectral types, with K and M dwarfs often displaying substantial photometric variability due to the presence of spots \citep[e.g.][]{sanchis-ojeda2011,mancini2017,roettenbacher2017}. This photometric variability can be used to determine the rotation period of the star. 

Active regions on stellar surfaces present a challenge to characterizing exoplanet atmospheres during transit, especially for K- and M-dwarf hosts. Unocculted spots and/or faculae introduce a spectrally varying offset into the inferred transit spectrum which can mimic planetary features, a phenomenon dubbed the Transit Light Source Effect \citep[TLSE;][]{rackham2018,rackham2019}. Depending on the spectral type, unocculted stellar features can mimic planetary absorption due to H$_2$O or CO, or masquerade as scattering hazes \citep[e.g.,][]{barstow2015,wakeford2019,may2023,Moran2023}. The TLSE can also change between different observations of the same target, creating challenges for combining transits \citep[e.g.][]{ahrer2025b}. Understanding the spectral signatures of spot and faculae contrasts is therefore critical for mitigating the TLSE in exoplanet atmosphere retrievals. 

Models of stellar active regions are becoming increasingly sophisticated,  \citep[e.g.,][]{norris2023,smitha2025}, but observational constraints on these features are relatively sparse, especially for faculae as their impact is greatest in the UV at wavelengths <500 nm (see Figure~\ref{figure:synthetics}). Spot crossings within planetary transits provide some direct information, allowing the size and temperature of the occulted spot to be determined from spectral lightcurve fitting \citep[e.g.][]{fourniertondreau2025} using tools such as SPOTROD \citep{beky2014}, PyTranSpot \citep{juvan2018} and fleck \citep{morris2020}. However, empirical constraints on the spot or faculae spectral contrast are a bigger concern, given the importance for accurately interpreting exoplanet transit spectra where the stellar host is active. 

Some studies, such as \cite{espinoza2019} and \cite{fourniertondreau2025}, present spot/photosphere contrast spectra extracted from lightcurve fits where spot crossings are present. \cite{fourniertondreau2025} show this for K dwarf WASP-52, and compare the extracted contrast spectra to PHOENIX stellar atmosphere models \citep{husser2013}. This allows an independent constraint to be placed on the starspot temperature relative to the photosphere, based on the PHOENIX model fits; however, there is remaining structure in the residuals for these PHOENIX model fits that could indicate that PHOENIX stellar models are not a good representation of star spot spectra. Indeed, \cite{smitha2025} find that spot contrasts for K and M dwarfs calculated using 1D radiative equilibrium models do not match contrasts generated using full 3D magnetohydrodynamic (MHD) simulations. 1D simulations fail particularly at wavelengths around 2 microns \citep[][their Figure 4]{smitha2025}, where the spot spectra extracted by \cite{fourniertondreau2025} are also not well represented by PHOENIX models. Similar discrepancies are found for faculae \citep[e.g.][]{witzke2022,norris2023}. In their Figure 9, \cite{norris2023} show that 3D MHD models for K0 stars produce stronger faculae contrasts in optical and near-ultraviolet wavelengths than equivalent 1D simulations, and also features at near infrared wavelengths that are absent for 1D models. 

A panchromatic spectrum of a polar spot on convective M dwarf TOI-3884 was obtained via a series of JWST observations using NIRISS and NIRSpec, spanning 0.6 -- 5.3 $\upmu$m, presented by \cite{murray2026}. They find that whilst 1D models provide a reasonable match to the observed spot spectrum at wavelengths longer than 1 $\upmu$m, at optical wavelengths the match is poorer. The authors point out that spot properties inferred only from optical wavelengths would likely be different to those inferred from infrared wavelengths, highlighting inconsistencies in 1D models and the need for broad wavelength coverage in observations of starspots. 

Transit observations of planets around active hosts provide one of very few opportunities to obtain observational data for starspot properties, with the aforementioned spot crossings, potential out of transit variation in spot coverage, and the transit itself each providing different pieces of information to complete the puzzle. The multiple JWST observations of the TRAPPIST-1 system's 7 planets are a good example, with flares observed in several observations and providing the opportunity for characterization \citep{lim2023,espinoza2025,vasilyev2025}, as well as observations of changing spot/faculae properties during different transits \citep{lim2023}. 

The precision and signal to noise of JWST, allowing transit spectra to be obtained from just a single transit event, has facilitated the drive to better understanding of stellar contamination. However, JWST is limited to wavelengths red of 0.6 microns, missing the region where spot and faculae contrasts are at their maximum, and in particular not covering key parts of the spectrum where 3D MHD faculae models deviate from 1D radiative equilibrium simulations. Whilst JWST is an extremely valuable observatory for studying starspots, especially via repeated transits around the same stellar host, shorter wavelength observations are likely to be key for empirically constraining MHD spot and faculae models.

The Hubble Space Telescope (HST) has been a workhorse observatory for over 35 years, covering the ultraviolet-optical spectrum as well as the near-infrared. It has been used extensively to observe exoplanet transits, especially using the Space Telescope Imaging Spectrograph (STIS) in the UV and optical, and Wide Field Camera 3 (WFC3) across UV-optical-infrared wavelengths \citep[see e.g.][for example]{sing2016}. Archival HST datasets provide a wealth of information about stellar heterogeneities, although the need to combine spectra taken over multiple epochs to cover the full spectral range complicates their interpretation \citep{niraula2026}. In more recent years, the UVIS G280 grism on WFC3 has been used to characterize the UV-optical transit of a number of exoplanets. This mode has the advantage of obtaining wavelengths from 0.2 to 0.8 microns in one go, although for the majority of targets two or more transits are required for sufficient signal-to-noise \citep[e.g][]{lothringer2022,wakeford2020,boehm2025,gascon2025}. 

Finally, in recent years retrieval analyses of exoplanet transit spectra have accounted for unocculted spots and faculae outside of the transit chord. Parametric representations of spots and faculae typically interpolate over PHOENIX model grids in temperature space, allowing the temperature contrast and coverage fraction to be constrained. This exercise has been performed for HST and JWST data \citep[e.g.][]{pinhas2018,rathcke2021,lim2023,fourniertondreau2025}. To an extent, the spot coverage fraction and temperature contrast are degenerate. There is also some degeneracy between the spot and faculae fractions when both are included, since both affect the spectrum across a broad range of wavelengths but in opposite directions, and for both the effect is strongest at shorter wavelengths. Where multiple transits exist for the same target or at least the same star \citep[e.g][]{lim2023,ahrer2025b}, retrieval analysis has revealed that the stellar activity signature is often time-variable; therefore, unocculted stellar heterogeneity signatures cannot reliably be extracted where transits are obtained at different times and either stitched or coadded to produce the final spectrum \citep{ahrer2025b}. 

Obtaining constraints on unocculted starspots across a broad wavelength range requires simultaneous observation due to the variability of spot and faculae coverage fractions. Broader wavelength coverage, particularly extending into the optical and UV, can help to disentangle degeneracies between coverage fraction and temperature contrast and also the contributions from spots and faculae. With current instrumentation, the only way to achieve this is with simultaneous observation using HST and JWST.

In this paper, we present synthetic transit spectroscopy observations of canonical hot Jupiter HD 189733 b \citep{bouchy2005}, and use these to demonstrate that simultaneous UV-optical-NIR observation across multiple transits is instrumental for empirically constraining the properties of spots and faculae. HD 189733 is a known active K2 dwarf \citep{boisse2009}. The existing HST transit spectrum of HD 189733 b represents the first example of a correction being applied to transit data to account for unocculted starspots \citep{pont2013}. In this instance, the unocculted starspot coverage fraction and subsequent correction were inferred from long term ground-based monitoring of the target, and this was applied separately to each transit observation before stitching together the datasets from different instruments. HD 189733 b also provides an example of a steep optical scattering slope in the planet's atmosphere, a phenomenon that is often found to be degenerate with unocculted starspot signatures. 

\section{Synthetic Observations}

\subsection{NEMESISPY}
We generate the synthetic observations and perform retrieval tests using the NEMESISPY retrieval code. NEMESISPY \citep{yang2024} is a python version of the NEMESIS \citep{irwin2008} retrieval package that has been optimized for analysing observations of transiting exoplanets. It combines a fast radiative transfer model, using correlated-k tables for molecular opacities \citep{lacis_oinas1991}, with the PyMultiNest nested sampling algorithm. NEMESISPY has been used to analyse spectra of multiple planets observed with both Hubble and JWST, for example KELT-7b \citep{gascon2025,ahrer2025}, WASP-43b \citep{yang2024b} and L98-59 d \citep{banerjee2024}. 

\begin{table}
\centering
\label{table:synthetics}
\caption{This table contains the input values for synthetic model parameters and also the prior ranges for each used in the retrieval tests. }
\begin{tabular}{l|c|c}
        \hline 
        Parameter & Input Value & Prior Range \\
        \hline
        log$_{10}$(H$_2$O) & -4.0 & $U$(-12,-0.5)\\
        log$_{10}$(CO$_2$) & -5.0 & $U$(-12,-0.5)\\
        log$_{10}$(Na) & -5.0 & $U$(-12,-0.5)\\
        log$_{10}$(K) & -7.0 & $U$(-12,-0.5)\\
        $T_{\mathrm{P}}$(K) & 1100 & $U$(500,2000)\\
        log$_{10}$($P_{\mathrm{ref}}$/bar) & -1.5 & $U$(-9,1) \\
        log$_{10}$($P_{\mathrm{cloud}}$/bar) & -1.2 & $U$(-9,1) \\
        scattering index & 10 & $U$(0,14) \\
        opacity scaling & 6.2 & $U$(-5,10) \\
        \hline
        $T_{\mathrm{phot}}$(K) & 5050 & $N$(5000,200) \\
        $\Delta T_{\mathrm{spot}}$(K) & -650 & $U$(-1500,-250) \\
        $f_{\mathrm{spot}}$ & 0.01 & $U$(0,0.3) \\
         & 0.03 & \\
         & 0.1 & \\
        $f_{\mathrm{fac}}$ & 0.05 & $U$(0,0.3) \\
         & 0.1 & \\
         & 0.2 & \\
         \hline     
\end{tabular}
\end{table}

\begin{figure*}
	\includegraphics[width=\textwidth]{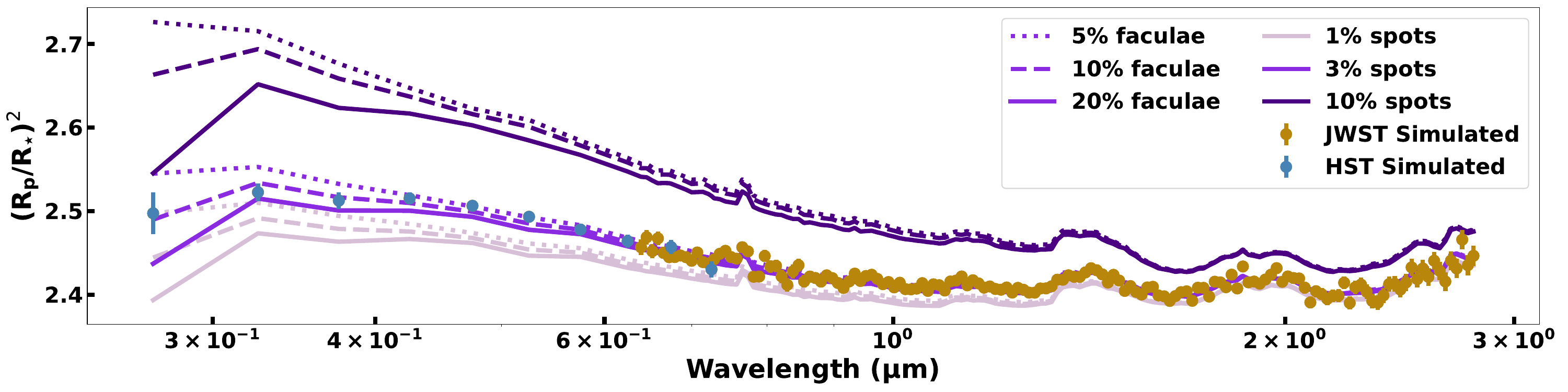}
    \caption{This figure shows the 3$\times$3 synthetic spectra generated from different combinations of starspot and faculae models. The shade of purple corresponds to the spot coverage fraction, with darker shades indicating more spots, whilst the linestyle denotes the facular coverage fraction, with more solid lines corresponding to higher facular coverage. The spectra are shown at the resolution of simulated Hubble WFC3/UVIS G280 and JWST NIRISS/SOSS observations that would have the appropriate signal-to-noise for a single transit observation. The gold and blue points show simulated data, including normally distributed noise, for the 10\% faculae, 3\% spots case. }
    \label{figure:synthetics}
\end{figure*}

\subsection{Model setup}
\label{models}
We set up our model atmosphere for the synthetic observations to be compatible with existing optical and near-infrared observations of HD 189733 b. We include opacity due to H$_2$O  \citep{Polyansky2018ExoMolWater}, CO$_2$  \citep{Yurchenko2020ExoMolCO2}, Na \citep{KURonline,allardsodium2019} and K \citep{KURonline,allardpotassium2016}. These species are selected since they are confirmed in HD 189733 b's atmosphere , and have strong absorption features at the wavelength of interest \citep{pont2013,fu2024}. We do not include CO and H$_2$S as they have weaker features in the NIRISS wavelength range and would therefore have a negligible effect on the results of this study. K tables are generated using the ExoMolOP database \citep{chubb2021}. We also include collision-induced absorption due to H$_2$ and He, parameterised according to \cite{borysow89,borysowfm89,borysow90,Borysow2001,borysow02}. 

We parameterize the cloud following the formalism of \cite{macdonald2017}, with an opaque grey cloud deck beneath a specified pressure, and a power law scattering haze layer above. We specify an isothermal temperature-pressure profile, since over the wavelength range of interest we do not expect to probe a large range of temperatures. 

Our stellar activity contamination model includes both spots and faculae. Spots and the unspotted stellar photosphere are modelled using PHOENIX \citep{husser2013} models of the appropriate temperature, via PYSYNPHOT \citep{pysynphot}. 
For the facular contrast we use the 300-G vs field-free K-star models presented in \cite{norris2023}. These are derived from MuRAM 'box-in-a-star' magnetoconvection simulations \citep[see, e.g.,][]{Voegler2005,beeck2013,Beeck2015b} that have been post-processed to calculate emergent intensities using ATLAS9 \citep{kurucz1993}. See \cite{norris2023}for a more detailed description of the facular contrasts. We take this approach for faculae since \cite{norris2023} and \cite{witzke2022} show that 1D faculae models do not capture the complex spectral and limb-dependent behaviour of the facular contrast. 

The stellar heterogeneity contamination $\epsilon_{\lambda}$is calculated as follows:
\begin{equation}
    \epsilon_{\lambda}= \frac{1}{
1 -
\left[f_{\mathrm{spot}} \left(
1 - \frac{I_{\lambda,\mathrm{spot}}(T_{\mathrm{spot}})} {I_{\lambda,\mathrm{phot}}(T_{\mathrm{phot}})}
\right) + f_{\mathrm{fac}} c_{\mathrm{fac}}
\right]
}
\end{equation}

where $f_{\mathrm{spot/fac}}$ are the fractions of the visible stellar disc covered by spots and faculae; $T_{\mathrm{spot/phot}}$ are the temperatures of the spots and unspotted photosphere; $I_{\lambda,{\mathrm{spot/phot}}}$ are the PHOENIX spectra of the spots and unspotted photosphere at the relevant temperatures; and $c_{\mathrm{fac}}$ is the facular contrast calculated from the MuRAM simulations.

We generate a 3$\times$3 grid of synthetic spectra with different spot and faculae coverage fractions. These spectra are designed to simulate the changes in spot and faculae coverage on the observable stellar disc during a stellar rotation period. Spot coverage varies between 1 and 10 \% whereas faculae coverage varies between 5 and 20\%. As facular contrasts are obtained from averages over the magnetoconvection simulation boxes they do not represent the contrast of individual small-scale flux tubes, but can be seen as 
average contrasts of an active region containing faculae. The facular filling factors quoted here are thus larger than, e.g., filling factors derived from solar magnetograms \citep[see][for further discussion]{johnson2021}.   
The input parameters for the synthetic spectra are listed in Table 1. We display the 9 generated synthetic spectra in Figure~\ref{figure:synthetics}. The faculae contrast is largest towards the shortest wavelengths, whilst the spot contrast persists into the near infrared.

\subsection{Noise model}

For the synthetic retrievals, we add noise to the datasets for each simulated instrument. It is important to note that the noise draw is different for each model instance, designed to make our simulations more realistic. For NIRISS, we use PandExo \citep{batalha2017} to calculate the noise for NIRISS/SOSS orders 1 and 2. For WFC3/UVIS, we use the \href{https://github.com/hrwakeford/HST_WFC3_UVIS_G280_sim}{UVIS G280 sim} tool \citep{wakeford2020}. We add noise to the simulated spectra prior to retrieval by generating an array of Gaussian random numbers with standard deviation equal to the simulated error bar, and adding this to the model. Perturbing the model spectra in this way makes the retrieval results more realistic. 

\section{Retrievals}
One of the biggest challenges in transmission spectrum retrievals around active targets is disentangling the spectral signatures of spot and faculae contamination from other spectral continuum features such as cloud scattering. However, we suggest that it is possible to use the time-varying nature of the spot and faculae contamination to advantage, and here we demonstrate this through a synthetic retrieval study. If the star is not a slow rotator, it is a reasonable assumption that, within a handful of stellar rotations, the spot and faculae characteristics and also the planet's atmospheric properties do not change; the only parameters likely to vary are the spot and faculae coverage fractions. We can therefore perform joint retrievals of two spectra taken at different phases of the star's rotation curve with the addition of only two further free parameters - the spot and faculae coverage during the second set of observations. 

\begin{figure*}
	\includegraphics[width=\textwidth]{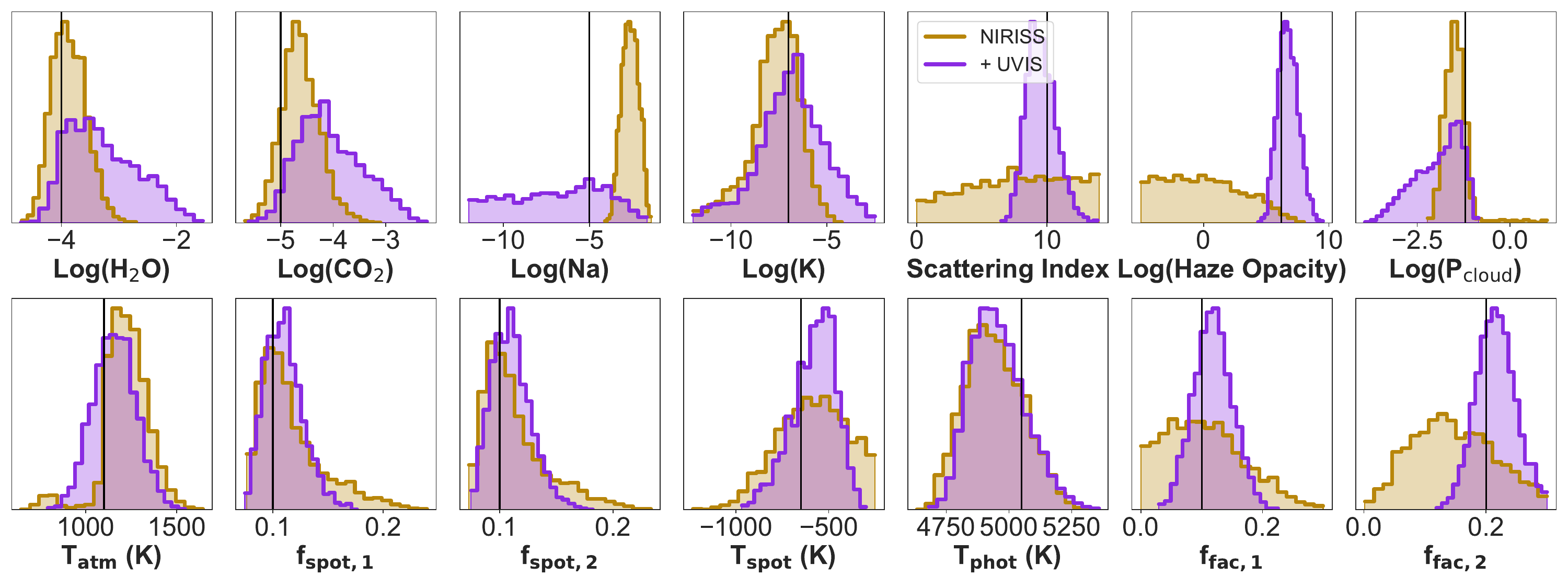}
    \caption{This figure shows a comparison across two synthetic retrievals, performed on spectra with [10\%, 10\%] spots and [10\%, 20\%] faculae respectively, with (purple) and without (gold) UVIS . The input values are indicated by vertical lines. The constraints on the haze properties, starspot temperature and starspot/faculae coverage fractions dramatically improve with the addition of the UVIS data (purple posteriors), whilst the constraints on the gas abundances and cloud top pressure become less precise.}
    \label{figure:69_uvis_niriss}
\end{figure*}

\begin{figure*}
	\includegraphics[width=\textwidth]{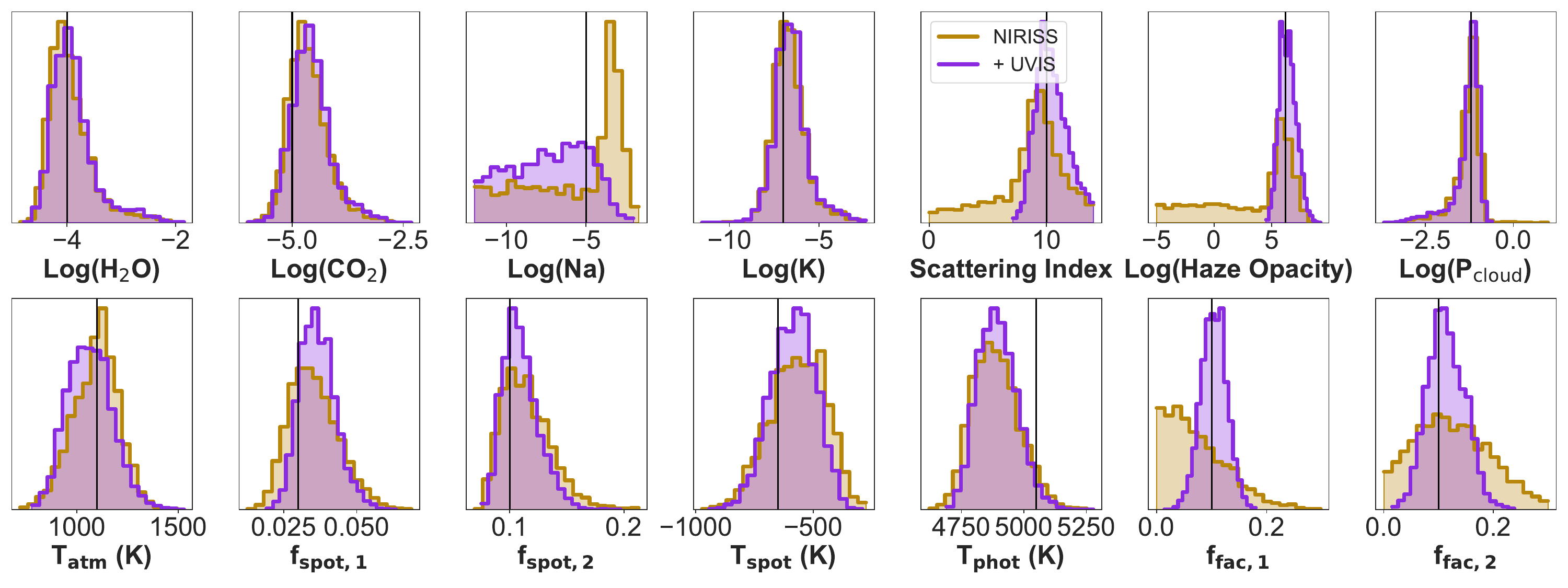}
    \caption{As Figure~\ref{figure:69_uvis_niriss}, performed on spectra with [3\%, 10\%] spots and [10\%, 10\%] faculae respectively.}
    \label{figure:56_uvis_niriss}
\end{figure*}

\begin{figure*}
	\includegraphics[width=\textwidth]{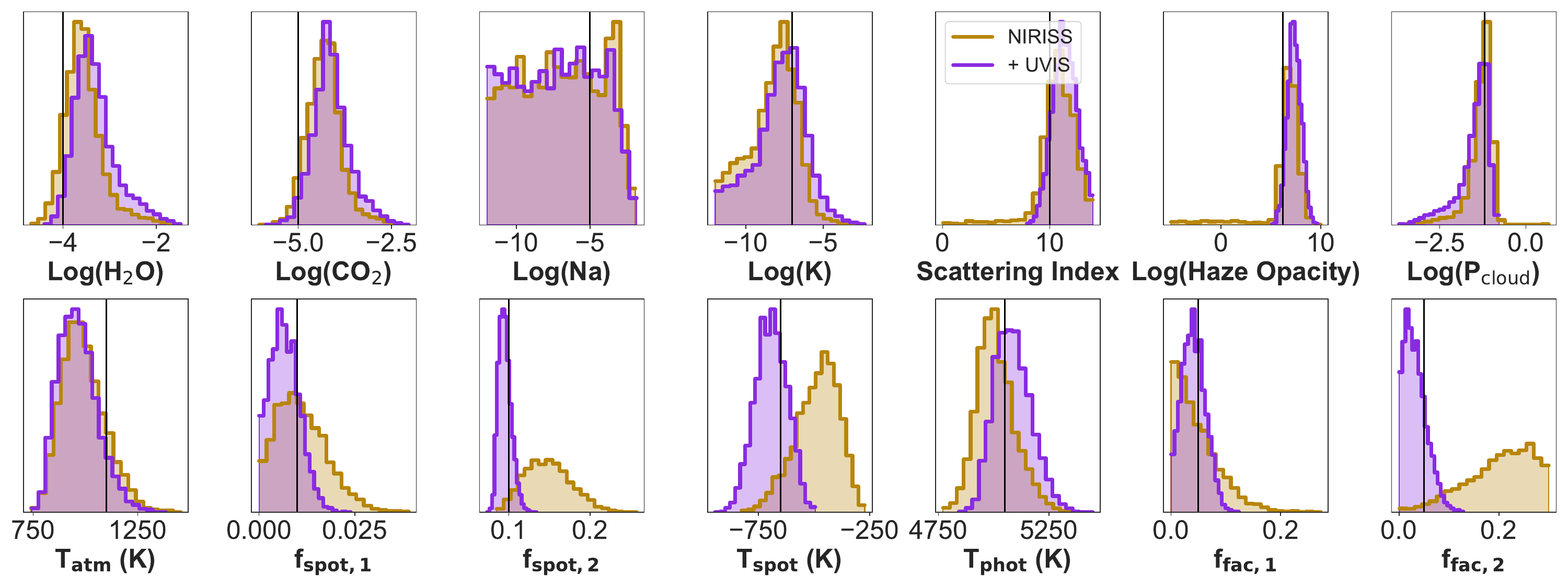}
    \caption{As Figure~\ref{figure:69_uvis_niriss}, performed on spectra with [1\%, 10\%] spots and [5\%, 5\%] faculae respectively.}
    \label{figure:13_uvis_niriss}
\end{figure*}

\begin{figure*}
	\includegraphics[width=\textwidth]{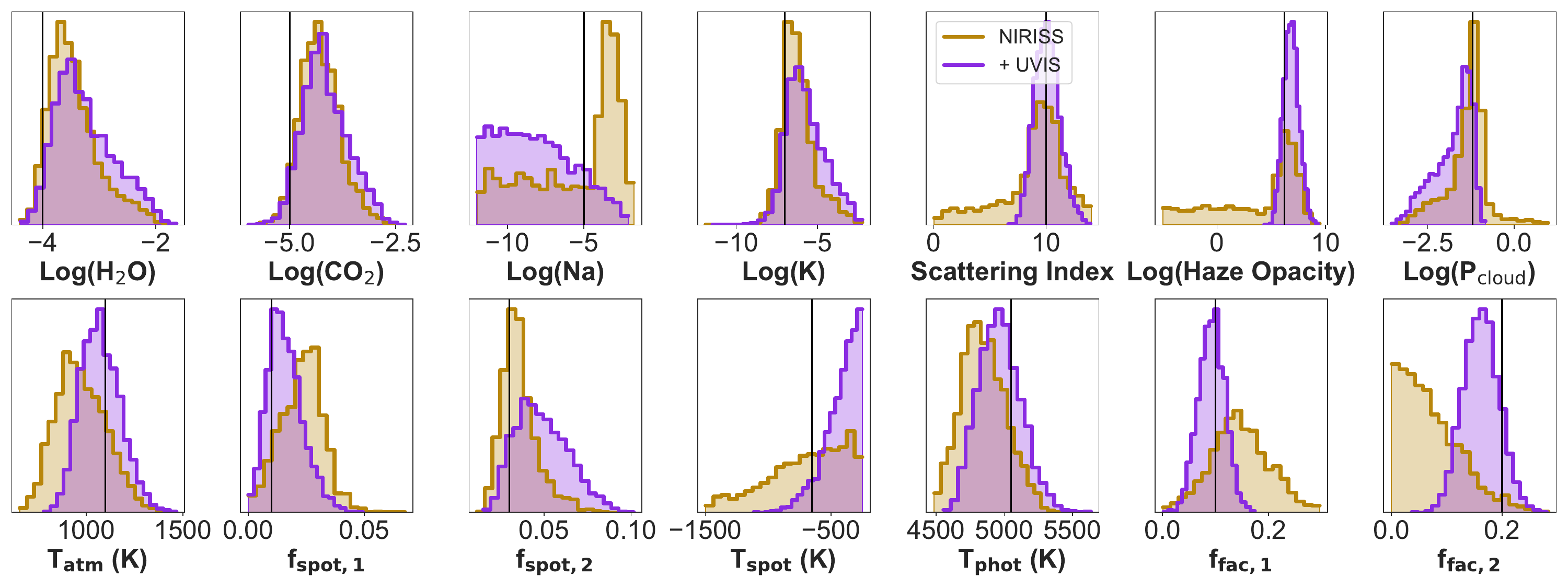}
    \caption{As Figure~\ref{figure:69_uvis_niriss}, performed on spectra with [1\%, 3\%] spots and [10\%, 20\%] faculae respectively.}
    \label{figure:48_uvis_niriss}
\end{figure*}

To simulate this effect, we perform joint retrievals on pairs of spectra which vary in spot and faculae coverage. We present examples in Figures~\ref{figure:69_uvis_niriss}, ~\ref{figure:56_uvis_niriss}, ~\ref{figure:13_uvis_niriss}, and ~\ref{figure:48_uvis_niriss}, showing different mixtures of spot and faculae coverage within each pair. In each figure, we display the results of a simulated retrieval from the NIRISS data alone and from the NIRISS data combined with data from UVIS G280. Comparing the retrieval results reveals the sensitivity of the result to data taken at different wavelengths. We show the posteriors for all retrieved parameters except the reference pressure, which is effectively a nuisance parameter accounting for our lack of prior knowledge of the pressure level for the white light planet radius. 

\subsection{The need for UVIS}
Here, we compare the retrievals with NIRISS SOSS alone with those combining SOSS and UVIS G280. We illustrate this for four different combinations of varying spot and facular coverage, using the following notation to denote the changing spot and faculae coverage: [spots$_1$, spots$_2$], [faculae$_1$, faculae$_2$]. The four different combinations are  [10\%, 10\%], [10\%, 20\%] (Figure~\ref{figure:69_uvis_niriss}); [3\%, 10\%], [10\%, 10\%] (Figure~\ref{figure:56_uvis_niriss}); [1\%, 10\%], [5\%, 5\%] (Figure~\ref{figure:13_uvis_niriss}); and [1\%, 3\%], [10\%, 20\%] (Figure~\ref{figure:48_uvis_niriss}). In these examples, we see the following consistent features emerging from the NIRISS-only retrievals relative to the retrievals including UVIS, which we discuss in more detail below:
\begin{itemize}
\item reduced ability to constrain starspot and faculae coverage fractions, especially faculae;
\item reduced ability to constrain aerosol properties (scattering index and haze opacity)
\item artificially enhanced precision on Na abundance
\end{itemize}
To a lesser extent, we also sometimws see an artificially enhanced precision on the H$_2$O abundance

We can see this in the example shown in Figure~\ref{figure:69_uvis_niriss}. The constraints on the starspot properties and haze properties improve substantially; in particular, with NIRISS only there is no constraint on the scattering index or haze opacity, and the faculae coverage fraction for the second spectrum would have been incorrectly retrieved. However, we can also see that the posteriors for the gas abundances and the cloud top pressure widen drastically; why does this happen?

If we examine the Log(Na) abundance, we see that for the NIRISS-only data we obtain a tight constraint, but that the retrieved abundance does not match the input. This is because the centre of the Na band is actually not covered by the NIRISS data, so the retrieved abundance depends only on the opacity from the line wings. The WFC3 UVIS G280 spectrum lacks the wavelength resolution to resolve the Na feature, so we have only an upper limit constraint on the abundance; however, this upper limit constraint is more correct than the artificially tight constraint from the NIRISS dataset alone. 

We also notice that the cloud top pressure Log(P$_{\mathrm{cloud}}$) constraint is tighter with NIRISS only. This is due to degeneracy between the cloud top pressure and the H$_2$O abundance, which in turn is affected by the constraints on Na.

Due to the broad Na wings, which extend under some of the H$_2$O features, the artificial precision on Na is likely to also produce a tighter constraint on H$_2$O. When the Na abundance posterior is wider, this sometimes corresponds to a wider H$_2$O posterior (Figure~\ref{figure:69_uvis_niriss} in particular). The extent to which parameters are degenerate in each example is influenced by the noise draw for that particular example, which is why we see some differences in behaviour between examples.  

\subsection{Sensitivity to faculae spectral contrast}
To test whether the retrieval is sensitive to the details of the faculae contrast spectrum, we perform a retrieval on one pair of spectra but for the retrieval model we use the faculae contrast taken from radiative equilibrium PHOENIX models rather than the contrasts derived from 3D magnetoconvection simulations that we used to generate the synthetic spectra. To keep all other aspects of the retrieval the same, we do not retrieve the faculae contrast temperature; instead, we use a fixed faculae contrast of +50 K for a photospheric temperature of 5050 K, which produces a similar overall magnitude of contrast for a given faculae coverage fraction; however, the shape of the contrast spectrum differs from the magnetoconvection model. 

\begin{figure*}
	\includegraphics[width=\textwidth]{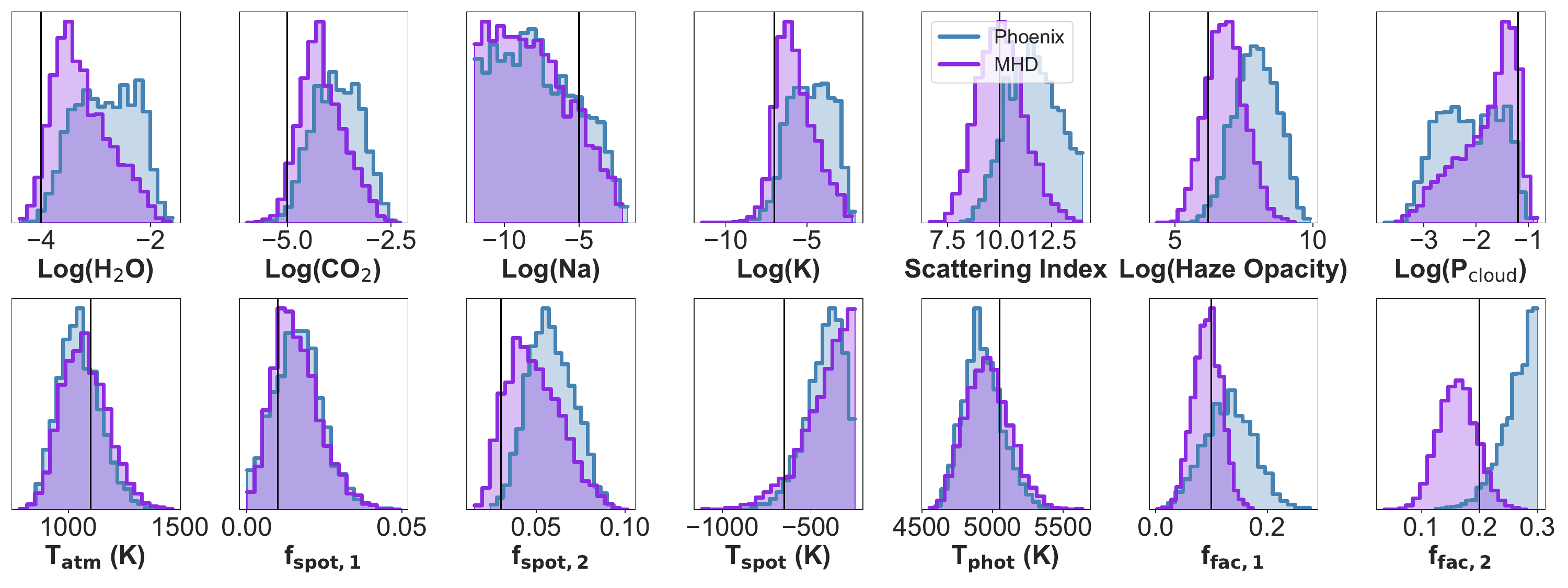}
    \caption{As Figure~\ref{figure:69_uvis_niriss}, but showing data for NIRISS+UVIS retrievals of spectra with [1\%, 3\%] spots and [10\%, 20\%] faculae respectively, comparing the retrievals using the PHOENIX-generated faculae contrast (teal) and the magnetoconvection contrast (purple). The retrieval results are biased from the true solution where the faculae contrast does not match the input.}
    \label{figure:48phoenix}
\end{figure*}

We find that retrieval results are biased when we use the PHOENIX-generated contrast, including the results for the planetary atmosphere (Figure~\ref{figure:48phoenix}). In particular, the H$_2$O abundance is overestimated and the cloud properties are incorrectly recovered, since it is in the ultraviolet region of the spectrum that this effect is most noticeable. The overall goodness of fit across both combined spectra is comparable between the PHOENIX and MHD model contrasts, with $\chi^2/n$ values of 0.870 and 0.864 for PHOENIX and 3D magnetoconvection respectively; however, in the UVIS G280 spectrum alone it is easier to distinguish, with $\chi^2/n$ values of 0.981 for the G280 spectrum only and 0.785 for the PHOENIX and MHD models. The UVIS G280 dataset therefore provides critical constraints that allow us to test and refine models of faculae contrast. 

\subsection{Co-adding spectra with unocculted heterogeneities}

Co-adding spectra obtained at different times is sometimes necessary where the signal-to-noise from a single observation is insufficient. This is likely to be required in particular for the Ariel mission \citep{tinetti2018,edwards2019} to achieve the required signal-to-noise for all mission tiers. Co-adding spectra for planets orbiting active stars is potentially problematic, since standard co-adding averages over two (or more) different spot/faculae configurations, and this may bias the retrieval result. 

We have shown that the combination of NUV-optical and near-infrared is required for appropriate characterization of the spots and faculae themselves, but for Ariel and for instances where coadding is required for JWST, there will be no NUV data available. If the signal to noise is such that coadding is required to recover the planetary atmosphere properties, there is also unlikely to be sufficient signal to constrain the spot and faculae contributions within individual spectra. The important question is, for red optical and near-infrared spectra, does a simple average over different spot/faculae configurations introduce bias in the retrieved planetary atmosphere properties? 

\begin{figure*}
	\includegraphics[width=\textwidth]{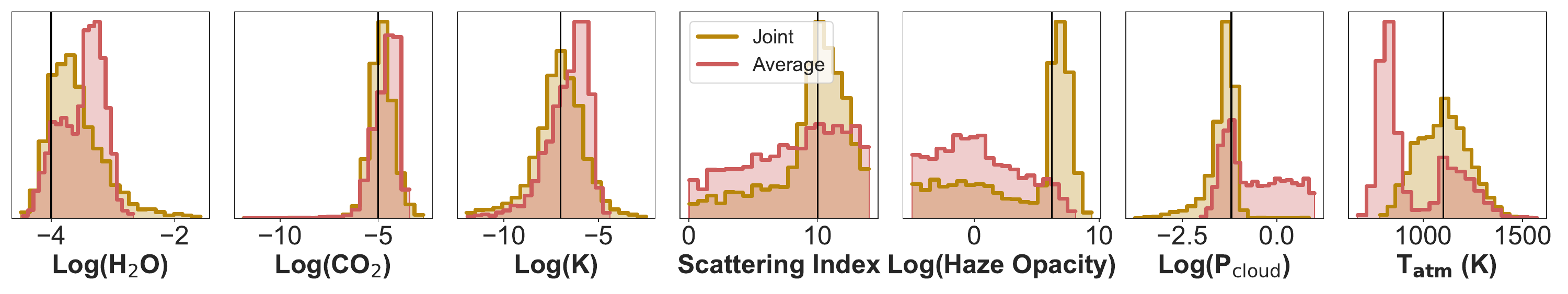}
    \caption{As Figure~\ref{figure:69_uvis_niriss}, but showing planetary data for the NIRISS-only joint retrieval of spectra with [10\%, 1\%] spots and [5\%, 20\%] faculae respectively, compared with a single retrieval on the average of those two spectra. We omit the Na posterior as the Na line centre is not covered by NIRISS.}
    \label{figure:37average}
\end{figure*}

Here, we test this by comparing our standard two-spectrum retrieval framework with a retrieval on the averaged spectrum. In figure~\ref{figure:37average}, we present a case where the retrieved planetary properties from the averaged spectrum do indeed show bias away from the input values. The averaged spectrum retrieval produces a multi-modal posterior, with the peak of the higher probability mode offset from the input values, whilst the joint retrieval more accurately recovers the atmospheric parameters. We still accounted for starspots and faculae in the averaged retrieval, but the retrieved results did not resemble the spot solution from either individual spectrum. 

In other cases that we tested, the solutions using the joint retrieval and those from the averaged retrieval were not significantly different. This is reassuring from the perspective of Ariel, since this mission will rely heavily on coadding for some targets. However, it is clear that the precise combination of spots and faculae dictates whether or not the average spectrum is a sufficiently good representation to extract the planetary parameters accurately. Importantly, it is not reasonable to assume that an averaged spectrum is always sufficient, and running joint retrievals of the kind shown here as well as a coadded retrieval is likely to be an important test where multiple spectra are required to achieve the desired signal-to-noise. 

\section{Discussion}
Through the synthetic retrieval tests above, we have shown that the changing spot and faculae coverage is an important consideration for free retrievals of a target resembling HD 189733b. Here, we discuss the limitations of this study and how the results can be applied to a broader range of cases. 

\subsection{Uniqueness of HD 189733}
The method presented in this paper relies on the signal-to-noise of a single observation being sufficient to distinguish different star spot and faculae signatures with both WFC3/UVIS G280 and NIRISS/SOSS. Whilst this is possible for a large range of planets with NIRISS, the achievable signal-to-noise from WFC3/UVIS G280 is substantially more limited, and HD 189733 is close to the limiting case for this instrument with a single transit. To maximise the Transit Light Source Effect signal, ideally the transiting planet should be a gas giant. A gas giant target is also preferable because the atmospheric signal in the transit spectrum is strong, which makes it easier to disentangle the atmospheric signatures from those due to stellar activity. The only other main-sequence star with a transiting gas giant and a V-band magnitude greater than or equivalent to HD 189733 (Vmag: 7.67) is HD 209458, which is known to be a relatively inactive star. Therefore, HD 189733 is the outstanding active stellar target for which it is currently possible to measure the Transit Light Source Effect simultaneously across the NUV-optical-NIR range.

\subsection{Application to other targets}\label{others}
If we were to expand the possible target sample to include stars with smaller transiting planets, then for K dwarfs specifically there are only two stars with a V-band magnitude brighter than, or comparable to, HD 189733. These are HD 219134, which has two transiting super Earths and a further 4 non-transiting planets, and TOI-5789, which has one transiting super Earth and a further three non-transiting planets. For TOI-5789, its S-index indicates relatively low activity. HD 219134 may be a suitable target for similar investigations in future, but due to the lack of transiting gas giants it is a less suitable target than HD 189733. There are also several G dwarfs suitable for this type of observation, including 55 Cancri which hosts a transiting super Earth. Whilst they may not be as active as HD 189733, all stars will have some level of surface heterogeneities, and characterizing these is no less important for lower activity stars. Extending to slightly fainter K dwarfs may also be possible, with the caveat that a single transit with WFC3 UVIS must yield sufficient signal to noise to distinguish spectral features of spots and faculae. This would open up the possibility to perform these measurements for other active K dwarfs with transiting giant planets, e.g. WASP-69 or HAT-P-11.

Although performing the same direct observations would not be possible for the vast majority of active G and K dwarf host stars, and would be especially challenging for M dwarf hosts since these stars are less bright in the V-band relative to infrared wavelengths, placing empirical constraints on stellar surface features for even a few such targets can drive the future development of physically motivated faculae and spot models. Improvements in these models will in turn improve the accuracy of stellar heterogeneity corrections for other targets, even without direct constraints from ultraviolet data. Transits provide a rare empirical window into the specific spectral signatures of stellar surface features, and as such they are useful even where the data that can be obtained are limited. Starspot constraints are particularly critical for M dwarfs hosting rocky planets, which are typically highly active but also require multiple transits to achieve the required signal-to-noise. \cite{kreidberg_stevenson2025} assert that stellar contamination is the biggest roadblock to precise transmission spectroscopy of these planets, and empirical study is an important step in addressing this problem. 

\subsection{The TLSE in the absence of ultraviolet data}
We have shown in this paper that directly constraining spot and faculae coverage, and the spectroscopic signatures of spots and faculae, depends on ultraviolet data. There is insufficient information in JWST data to accurately constrain these properties. Atmospheric parameters retrieved at longer wavelengths are relatively unaffected by the accuracy of the retrieved TLSE parameters, at least in spectral regions where starspots do not introduce additional molecular absorption. We have demonstrated that whilst cloud properties are often poorly recovered with NIRISS data alone, gas abundances such as H$_2$O and CO$_2$ may still be recovered for K dwarf hosts even in the absence of accurate constraints on the TLSE parameters. We stress that this is not necessarily universal; e.g, for M dwarfs H$_2$O absorption in starspot spectra becomes critically important and can bias retrieved planetary H$_2$O abundance, see \cite{barstow2015}. It also does not mean that TLSE corrections do not need to be included in retrieval, rather that due to degeneracies inherent within the TLSE parameterization that TLSE corrections that do not accurately represent the true state of the stellar surface may still adequately remove the TLSE contamination from the planet spectrum. 

\section{Conclusions}
In this paper, we have demonstrated an observational scenario that would allow accurate characterization of the changing spot and faculae coverage for an active K dwarf target, and a benchmark dataset to refine models of stellar surface features. We have shown that the ultraviolet component of such a dataset would also be sensitive to the spectral shape of faculae contrast, which is much less apparent in infrared datasets. Although the targets that could be used to make this measurement with existing telescopes are very few, as discussed in Section~\ref{others}, we suggest that such observations would provide valuable empirical constraints on 3D magnetoconvection models for spots and faculae. 

We find that, when a retrieval is carried out assuming facular contrast from a 1D model on data generated using 3D magnetohydrodynamic simulations, bias may be introduced into the retrieval results that follow. This is critical, since it means that accurate representations of spot and faculae properties are important for properly interpreting transmission spectra affected by stellar contamination. Conversely, it also means that such datasets are an important testbed for improving these representations.

We find that although NIRISS data alone do not allow for the accurate recovery of spot and faculae coverage, the planetary gas abundance parameters are generally still well retrieved. The TLSE parameters for coverage fraction and contrast for each type of heterogeneity are somewhat degenerate with each other, and the degenerate solution often performs adequately as a correction for the atmospheric parameters. It does not however allow for accurate characterization of the stellar surface heterogeneities, and for this UV-optical data is necessary. We also stress that it is important to test the robustness of the corrections at NIRISS wavelengths against real datasets with full, simultaneous UV-optical-IR coverage, a resource that does not currently exist. 

Finally, we consider the case in which multiple infrared spectra with different TLSE characteristics are coadded to produce a spectrum with sufficient signal to noise to accurately recover the planet's atmospheric features. We demonstrate that in some cases, a simple coadding without accounting for potential differences in spot and faculae coverage between observations may bias the temperature, cloud and gas abundance values. We therefore recommend a joint fit to both spectra allowing for varying spot and faculae coverage as the more robust approach. This adds minimal computational expense whilst still combining the signals from both spectra to infer the planetary parameters.  

\section*{Acknowledgements}
JKB is supported by a UKRI Science and Technology Facilities Council Ernest Rutherford Fellowship (ST/T004479/1). CC is supported by a UKRI Science and Technology Facilitites Council PhD studentship. YCU acknowledges support from STFC Grant ST/W000989/1. This work used the DiRAC Data Intensive service (DIaL3) at the University of Leicester, managed by the University of Leicester Research Computing Service on behalf of the STFC DiRAC HPC Facility (projects dp320 and dp448).

\section*{Data Availability}
We do not make use of any observational data in this article. We make our synthetic observations, generated using the NEMESISPY code, available in supplementary online material. NEMESISPY is an open-source retrieval framework, available to install via pip (\href{https://pypi.org/project/nemesispy/}{NEMESISPY}). We use \href{https://pysynphot.readthedocs.io/en/latest/}{PYSYNPHOT} libraries to obtain the synthetic stellar photosphere and starspot spectra, and also for the 1D faculae models. Facular contrasts are taken from \citep{norris2023} and are available upon reasonable request from YCU.



\bibliographystyle{rasti}
\bibliography{example} 








\bsp	
\label{lastpage}
\end{document}